\documentclass[a4paper]{jpconf}
\usepackage{graphicx}
\usepackage{amsmath}
\usepackage{subfigure}
\usepackage{textcomp}
\usepackage[misc]{ifsym} 
\usepackage{hyperref}
\usepackage{threeparttable}
\usepackage{hyperref}
\usepackage{tikz}
\hypersetup
{
 colorlinks=true,
 linkcolor=blue,
 filecolor=blue,
 urlcolor=blue,
 citecolor=blue,
}
\usepackage[numbers]{natbib}  
\begin{document}
\begin{tikzpicture}[remember picture, overlay]
    \node[anchor=north west, align=left, font=\small] at ([yshift=-1cm]current page.north west) {%
        Submitted for publication in \\ 
        Journal of Physics: Conference Series \\ 
        (HAPP 10th Anniversary Commemorative Volume)
    };
\end{tikzpicture}

\title{SETI at FAST in China}

\author{
  \textsuperscript{1,2}Tong-Jie Zhang,
  \textsuperscript{1,2}Bo-Lun Huang, 
  \textsuperscript{1,2}Jian-Kang Li,
  \textsuperscript{1,2,3}Zhen-Zhao Tao,
  \textsuperscript{1,2}Xiao-Hang Luan,
  \textsuperscript{6}Zhi-Song Zhang,
  \textsuperscript{1,4,5}Yu-Chen Wang
}

\address{
  \textsuperscript{1}Institute for Frontiers in Astronomy and Astrophysics, Beijing Normal University, Beijing 102206, China \\
  \textsuperscript{2}Department of Astronomy, Beijing Normal University, Beijing 100875, China\\
  \textsuperscript{3}Institute for Astronomical Science, Dezhou University, Dezhou 253023, China\\
  \textsuperscript{4}Kavli Institute for Astronomy and Astrophysics, Peking University, Beijing 100871, China\\
  \textsuperscript{5}Department of Astronomy, School of Physics, Peking University, Beijing 100871, China\\
  \textsuperscript{6}National Astronomical Observatories, Chinese Academy of Sciences, Beijing 100012, China\\
  }

\ead{tjzhang@bnu.edu.cn}

\begin{abstract}
Since the commencement of the first SETI observation in 2019, China's Search for Extraterrestrial Intelligence program has garnered momentum through domestic support and international collaborations. Several observations targeting exoplanets and nearby stars have been conducted with the FAST. In 2023, the introduction of the Far Neighbour Project(FNP) marks a substantial leap forward, driven by the remarkable sensitivity of the FAST telescope and some of the novel observational techniques. The FNP seeks to methodically detect technosignatures from celestial bodies, including nearby stars, exoplanetary systems, Milky Way globular clusters, and more. This paper provides an overview of the progress achieved by SETI in China and offers insights into the distinct phases comprising the FNP. Additionally, it underscores the significance of this project's advancement and its potential contributions to the field.
\end{abstract}

\section{Introduction}

\begin{figure}[htb]
\centering
\subfigure{
\includegraphics[width=0.78\linewidth]{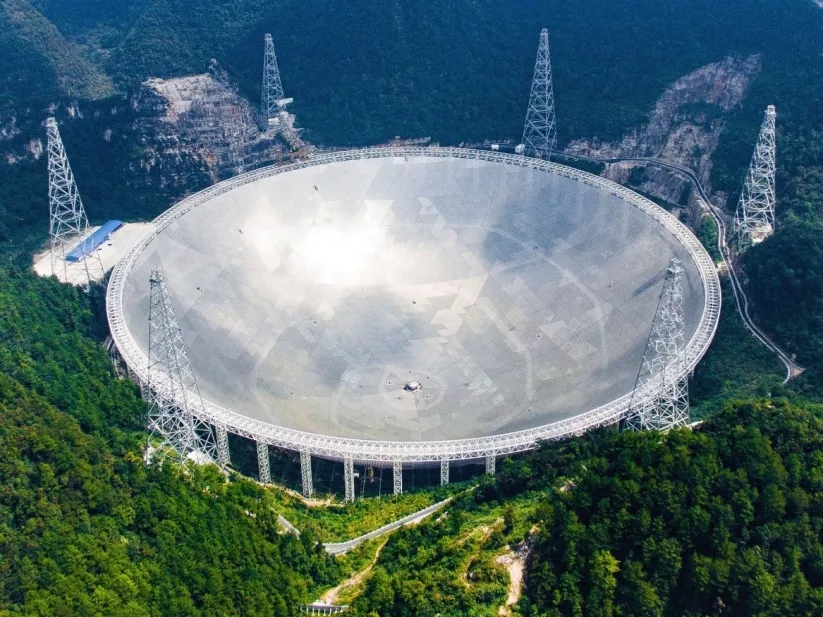}
}

\caption{A panoramic view of the FAST in early 2020 \cite{fastphoto}.}
\label{fig:res1}
\end{figure}

Humans have been wondering about the universe even before the emergence of our civilizations. With the belief that intelligent life inhabited among the stars would share similar ambitions and imaginations as we do, the concept of Search for Extraterrestrial Intelligence(SETI) was therefore raised in the early 20th century. In the past century, with the innovation and construction of radio telescopes, astronomers can finally look for artificial signals from outer space in the radio domain of the electromagnetic spectrum. Early SETI projects like Project Ozma and Project Phoenix laid the groundwork for subsequent efforts in the ongoing quest to detect signals or signs of intelligent life beyond Earth \cite{ozma, phoenix}. While no definitive evidence has been found to date, these endeavours have contributed to advancing our understanding of the cosmos and the potential existence of other intelligence. In 2016, The Five-hundred-meter Aperture Spherical Telescope(FAST), as shown in Fig.\ref{fig:res1}, started its collaboration with the Breakthrough Listen Initiative to search for weak radio technosignaures from nearby stars, and the first SETI observation in China was conducted in the year 2019. From 2019 to 2023, we have conducted multiple observations targeting various celestial objects including nearby stars, exoplanets found by the Kepler space telescope, and sky regions that are considered to contain potential extraterrestrial intelligence(ETI) sources by the SETI@home program. In 2023, with developments in data analysis techniques and observation strategies, we decided to propose the first long-term SETI project in China named the Far Neighbour Project(FNP) that aims to systematically search for ETI signals or leakage signals with a comprehensive workflow.

\section{SETI in China 2019-2023}
SETI is one of the five core scientific objectives of FAST, and its implementation commenced promptly following the completion of the FAST construction in 2016. In 2019, FAST was able to make its first step in the SETI observation, and multiple observations have taken place subsequently in the following years in collaboration with the Breakthrough Listen initiative and the University of California, Berkeley \cite{firstcn}.

\subsection{First SETI observation in China}
After three years of testing and technical adjustments, the development of compatible software and the installation of hardware, the first SETI observation in China was successfully conducted with the newly installed 19-beam receiver and the SERENDIP VI spectrometer on FAST \cite{Zhang2020jiyu}. With improvement in sensitivity, FAST is also receiving an enormous amount of weak radio frequency interference(RFI). Machine learning techniques were used to mitigate the RFI collected by FAST. This observation served as a test observation and it demonstrated that the majority of RFI was filtered out by the machine learning algorithm and the signal ranking pipeline called Nebula is also compatible with the observational data generated from FAST. The first SETI observation in China and its data analysis pipeline have proven the feasibility of SETI research with FAST, moreover, it marks the starting point of SETI in China \cite{firstcn}.

\subsection{Observation towards 33 exoplanets}
\begin{figure}[htb]
\centering
\subfigure{
\includegraphics[width=0.78\linewidth]{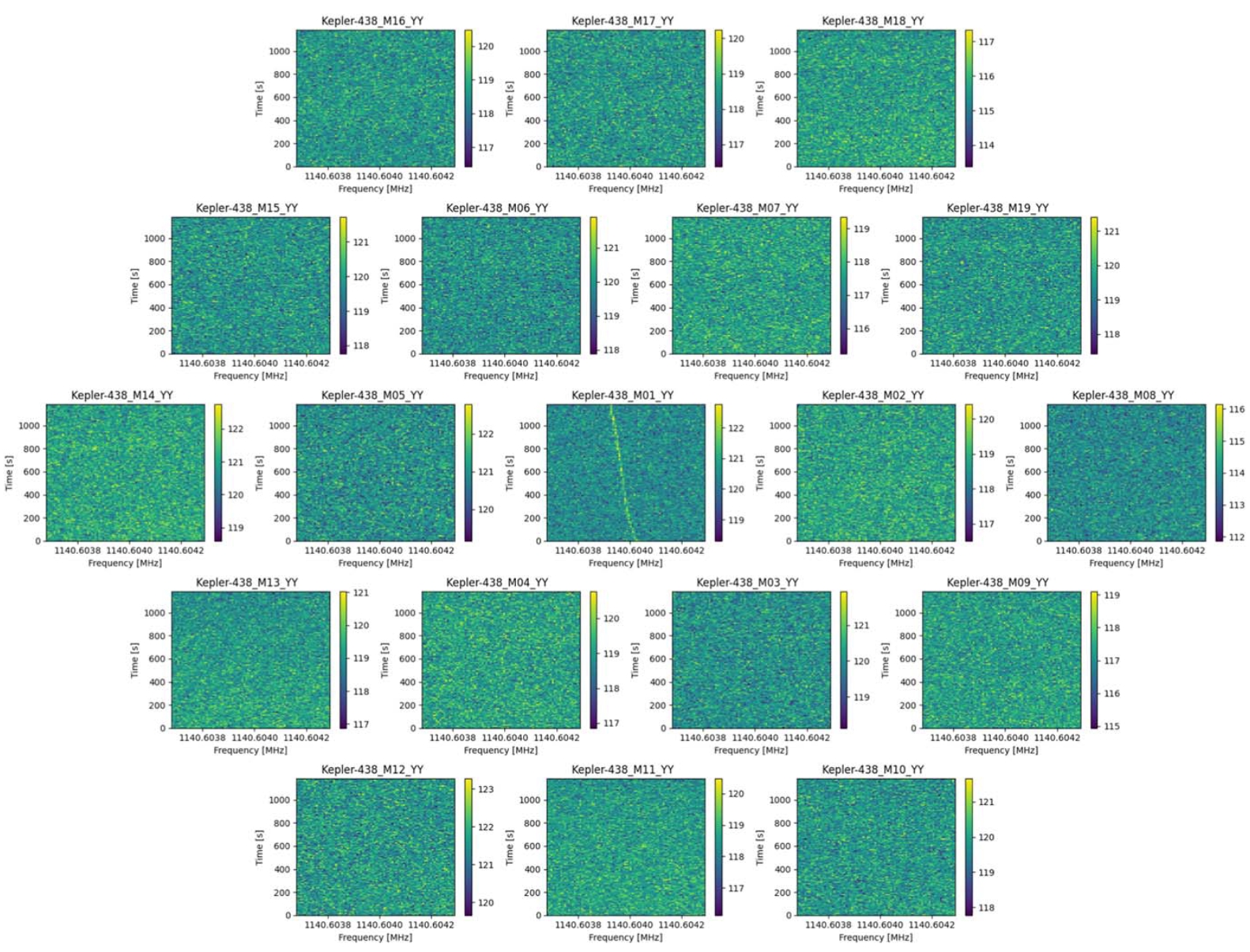}
}

\caption{The signal NBS 210629. It was detected around 1140.604 MHz during our 20-minute observation of Kepler-438. The relative position of the subfigures stands as a representation of the 19-beam receiver of the FAST. The signal is detected only in the central beam, and not in any of the reference beams.\cite{tao2022sensitive} }
\label{fig:resp}
\end{figure}

During April 2021 and September 2021, we conducted 11 rounds of targeted observations towards 33 exoplanetary systems \cite{tao2022sensitive}. These exoplanets were selected generally based on their habitability and whether or not they are in the Earth Transit Zone. In SETI, we usually quantify the sensitivity of a particular survey by the minimum detectable Effective Isotropic Radiated Power(EIRP) of a hypothetical extraterrestrial transmitter on that target. For the nearest exoplanetary system in this survey and its host star, Ross 128, the minimum detectable EIRP of a transmitter located in the Ross 128 stellar system for FAST is calculated to be $1.48\times 10^9$ W which is almost three orders of magnitudes less than the EIRP of the planetary radar mounted at the Arecibo observatory. Therefore, under the assumption that some of the alien civilizations are actively broadcasting their messages, FAST is capable of detecting such a signal even if the transmitter of the sender civilization possesses a significantly lower EIRP than the planetary radar at Arecibo. Furthermore, a new observation method is proposed and applied for this mission which is called the MultiBeam Coincidence Matching(MBCM) method. This method is similar to the ON-OFF method which has been widely utilized in the field of targeted SETI observation and the principle is that a sky-localized signal would not be detected in different telescope pointings(pointings are at least six half-power beam-width away from each other). The MBCM method takes advantage of the FAST's 19-beam receiver by simultaneously recording data in the centre beam and the six outermost beams. With the assistance of the MBCM method, we successfully eliminated all but one signal named NBS 210629 as can be seen in Fig. \ref{fig:resp}. However, it was later determined to be RFI by comparisons with RFI that showed similar frequency and polarization characteristics. In the re-observation towards NBS 210629 in November 2021, no signal that possesses similar characteristics was detected by the same instrument and observation strategy.

\vspace{\baselineskip}

\noindent In 2022, the observational data was revisited with the blind search mode of the MBCM method which aims to search for signals from the vicinity of the target exoplanets (radar or leakage radiation from a spacecraft) \cite{luan2023multibeam}. The blind search mode successfully found the signal NBS 210629, and another signal named NBS 210421 that the standard MBCM failed to return. However, NBS 210421 was thought to most likely come from a ground-based RFI due to similar features with the other four obvious RFI.

\subsection{Observation targeting nearby stars}

\begin{figure}[htb]
\centering
\subfigure{
\includegraphics[width=0.78\linewidth]{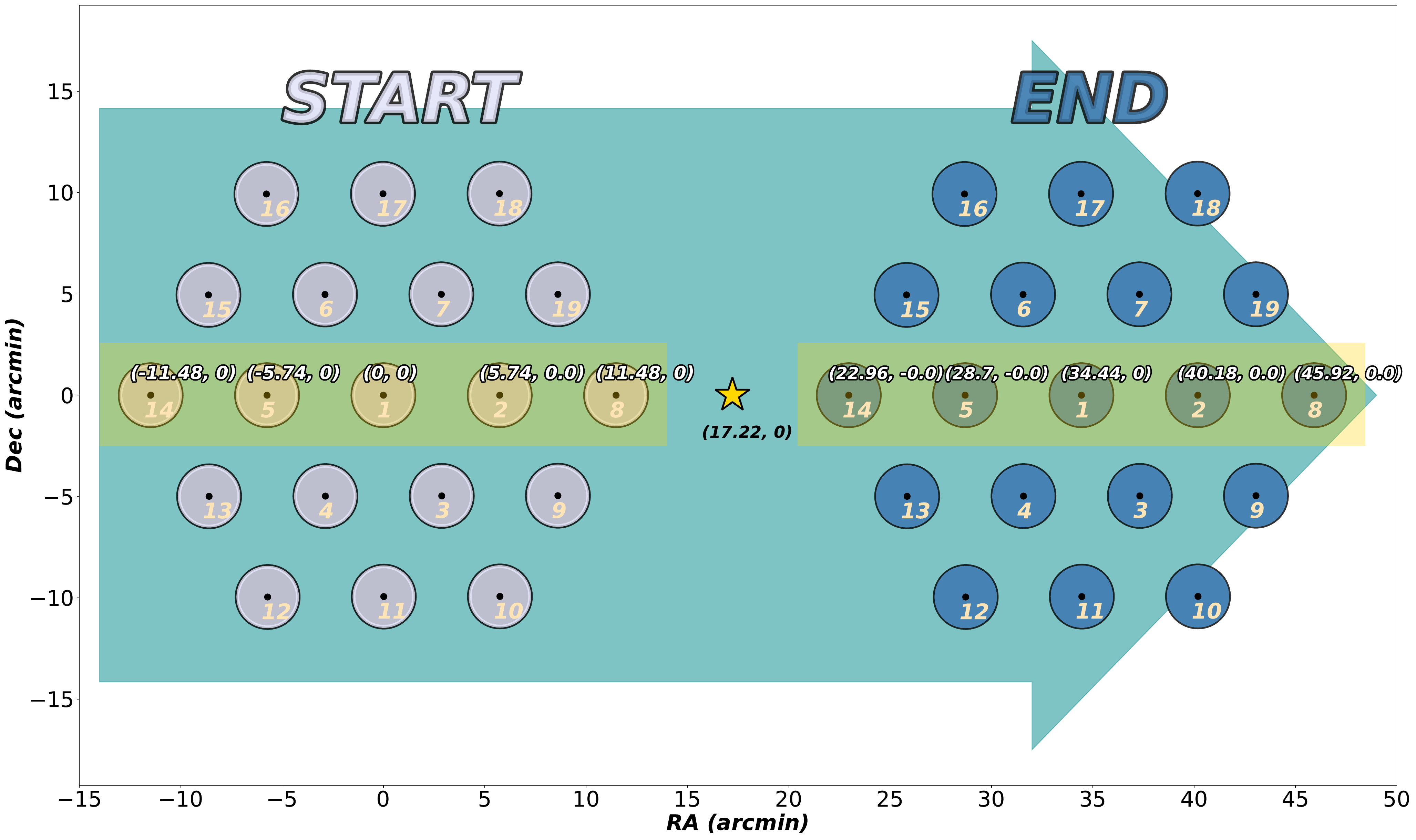}
}

\caption{The MBPS works by scanning the target object, each of the 19 beams would show a different representation of the signal and it is time-dependent because of their slightly different pointings during the scanning observation. \cite{huang2023solution}}
\label{fig:res3}
\end{figure}

Nearby stars have been a popular type of observation target since the beginning of SETI research. Suppose we are under the assumption that the distribution or model that describes the possible range of power levels of signals emitted by extraterrestrial transmitters follows a power law, which means that lower power transmitters are more abundant than those with a higher power. FAST, as the world's largest filled-aperture radio telescope, is ideal for searching weak radio signals from nearby stars with its extreme sensitivity. In 2022, we observed 14 nearby stars which were selected by their distance and whether they were in the Earth Transit Zone. However, no signal was considered a candidate ETI signal. In 2023, we conducted an observation of Barnard's star system with the minimum detectable EIRP reaching $4.36\times 10^8$ which is impossible to achieve by any other instrument to date \citep{TAO2023}. Furthermore, we developed a novel observation strategy called MultiBeam Point-source scanning(MBPS) observation as shown in Fig.\ref{fig:res3} based on the configuration of the FAST's 19-beam receiver and tested it in an observation towards the TRAPPIST-1 system. The MBPS is especially sensitive to persistent narrowband signal detection because a cross-verification procedure can be implemented with several newly introduced parameters in a single observation \cite{huang2023solution}.

\subsection{Observation with SETI@home}

With our collaboration with the SETI@home program, we re-observed the most promising candidates from the SETI@home Arecibo Sky survey using FAST. The SETI@home team has collected a list of the top SETI candidate sky regions obtained from 14 years of commensal multibeam observation at Arecibo \cite{werthimer2001berkeley}. This data collection involved numerous years of volunteer computing time, followed by analysis using the Nebula SETI data pipeline. In 2022, we conducted an observation of 51 sky regions from the SETI@home candidate list, and we are currently in the process of developing a new algorithm to search for signals that were previously detected by the SETI@home Arecibo sky survey. 

\vspace{\baselineskip}

\noindent To achieve an RFI excision like in the Nebula pipeline, we introduced a method involving the detection of continuous narrowband RFI using a threshold for sky separation and identifying drifting RFI using the Hough transform. Additionally, we incorporate machine learning to further eliminate RFI and pinpoint potential candidates. Our study compares this approach to previous work on the same FAST data. Despite its computational simplicity, the new method efficiently removed more RFI while retaining simulated ETI signals, except for those severely affected by RFI. Moreover, we identified a greater number of candidate signals, encompassing about twelve new candidates not previously reported \cite{wang2023search}.

\section{Far Neighbour Project} 
With increasing observation opportunities at FAST and the desire to explore the unknowns, we initiated the Far Neighbour Project(FNP) at Beijing Normal University in the summer of 2023. The FNP stands as a long-term project, there are generally four stages in this project.

\subsection{Stage I: Search for technosignaures beyond Earth}

At the initial stage of the FNP, we aim to systematically search for technosignaures from various celestial objects under a rigorous framework and pipeline. The FNP target selection strategy is designed in a way that maximizes the probability of detecting ETI signals. For example, we will conduct observation towards some of the Milky Way globular clusters as well as nearby stars in the first year of the FNP. The list of observation targets is updated every year, and it is determined according to the latest SETI theories and other related fields like astrobiology, exoplanetary science, etc. With our collaborative relationship with the Breakthrough Listen initiative and the SETI@home program, we continue the re-observation to observe the best candidates from the SETI@home Arecibo Sky survey. And together with the Breakthrough Listen team, we test and develop new data analysis techniques to remove RFI like the polarization characteristics, and dispersion feature \cite{li2023polarization, gajjar2022searching,li2022drift}. 

\subsection{Stage II: Searching and messaging}
This stage is initiated only if more than one high-confidence ETI signal is found. In this stage, the FNP will endeavour to decode the incoming messages and conduct detailed investigations of the source objects. Once an agreement is established based on the results of the investigation, a simple and straightforward message will be transmitted to the source object from a spacecraft that is located at a sufficiently distant region from Earth, the exact distance is determined by how far the source object is and the advancement of our technology at that time. We expect that the FNP will remain at this stage for a timescale of tens of years (messages from nearby stars), thousands of years (messages from Milky Way globular clusters), or longer than millions of years (messages from other galaxies).

\subsection{Stage III: Searching and Communication}

This stage will be heavily dependent on our technological and cultural development at that time, thus it is difficult to even broadly predict the timescale. Generally, in this stage, we will progressively establish trust through consistent and cautious communication and explore limited collaborations in areas like information exchange or basic scientific knowledge.

\subsection{Stage IV: The Far Neighbour atlas }

This stage will commence after we have gained a comprehensive understanding of multiple extraterrestrial civilizations. By working with various civilizations, the FNP will be assisting in the writing of the Far Neighbour atlas, both a map and catalogue that describe the vastness of the Milky Way and more importantly, the civilizations that reside within it.

\section{Conclusion}
To this date, our species has been conducting SETI research under a series of scientific theories for decades which is merely a tiny fraction even compared to the age of our civilization. With China's involvement in SETI research, we will accelerate the progress of SETI studies and conduct more sensitive observations of a greater number of celestial targets. Our novel observation methods, such as MBCM and MBPS, are expected to significantly enhance the efficiency and accuracy of data processing. In addition to observational efforts, we are also committed to integrating multiple disciplines to continuously refine the theoretical frontier of SETI.

\vspace{\baselineskip}

\noindent  In the FNP, we are dealing with signals with unknown characteristics, therefore, we propose the FNP as a long-term project without a predetermined duration. For Stage I, observations and development of SETI theories and new data processing techniques are the major focus. However, in this paper, the descriptions for Stages II to IV are relatively succinct compared to Stage I. This is because we envision the FNP as a program that can span over a millennium and more, provided our civilization endures. Given such an extended timescale, the rapid pace of technological development and scientific discoveries might significantly advance the methods and efficiencies outlined in our project. Therefore, we have refrained from elaborately detailing the implementation of stages II to IV, which have yet to be initiated. To date, we possess no evidence pertaining to the existence of extraterrestrial civilizations, let alone their spatial distribution. Consequently, the prospective outcome of the FNP remains challenging to anticipate. Nonetheless, even following meticulous observation and analysis, should the quest for any discernible ETI signals yield no results, we shall attain a more profound comprehension of the distribution of technological civilizations in the cosmos. Additionally, while conducting SETI research, we will also use data from SETI observations to explore other auxiliary scientific research in the field of astrophysics. 

\vspace{\baselineskip}

\noindent

\section*{Acknowledgments}
This work was supported by the National SKA Program of China (2022SKA0110202) and the National Natural Science Foundation of China (grants No. 11929301).

\bibliography{fnpref}

\end{document}